# Large magnetoresistance in an electric field controlled antiferromagnetic tunnel junction

Yurong Su[1], Jia Zhang[2*], Jing-Tao Lü[2], Jeongmin Hong[1], and Long You[1*]

[1]*School of Optical and Electronic Information, Huazhong University of Science and Technology, 430074 Wuhan, China*

[2]*School of Physics and Wuhan National High Magnetic Field Center, Huazhong University of Science and Technology, 430074 Wuhan, China*

[*]jiazhang@hust.edu.cn;

[*]lyou@hust.edu.cn.

Large magnetoresistance effect controlled by electric field rather than magnetic field or electric current is a preferable routine for designing low power consumption magnetoresistance-based spintronic devices. Here we propose an electric-field controlled antiferromagnetic (AFM) tunnel junction with structure of piezoelectric substrate/$Mn_3Pt$/$SrTiO_3$/Pt operating by the magnetic phase transition (MPT) of antiferromagnet $Mn_3Pt$ through its magneto-volume effect. The transport properties of the proposed AFM tunnel junction have been investigated by employing first-principles calculations. Our results show that a magnetoresistance over hundreds of percent is achievable when $Mn_3Pt$ undergoes MPT from a collinear AFM state to a non-collinear AFM state. Band structure analysis based on density functional calculations shows that the large TMR can be attributed to the joint effect of significant different Fermi surface of $Mn_3Pt$ at two AFM phases and the band symmetry filtering effect of the $SrTiO_3$ tunnel barrier. In addition, other than single-crystalline tunnel barrier, we also discuss the robustness of the proposed magnetoresistance effect by considering amorphous $AlO_x$ barrier. Our results may open perspective way for effectively electrical writing and reading of the AFM state and its application in energy efficient magnetic memory devices.

## I. INTRODUCTION

A conventional magnetic tunnel junction (MTJ) with high tunnel magnetoresistance (TMR) consists of two ferromagnetic (FM) electrodes separated by an ultrathin insulating barrier. The TMR effect can be applied in various spintronic devices, for instance, magnetoresistive random access memories (MRAM), magnetic field sensors, read heads for hard drives, and spin logics [1-3]. At current stage, electric current has been mainly used to manipulate the magnetization states in MTJs via spin-transfer-torque (STT) [4-6], spin-orbit torque [7-9] mechanisms or magnetic fields produced by current, which limits the energy efficiency of MTJ-based spintronic devices. Therefore, there have been great efforts aiming at electric field control of magnetic states instead of electric current. Voltage controlled magnetic anisotropy (VCMA) in perpendicularly magnetized CoFe(B)/MgO interface is one of the promising strategy [10-13]. However, the VCMA effect mainly relies on the electrostatic screening of magnetic electrodes at interface, and the voltage controlled effect alone is not sufficient for effectively manipulating magnetization state [10]. Alternative strategy towards electric-field control of TMR is to use ferroelectric barrier in a tunnel junction [14-16]. However, one of the main disadvantages of the ferroelectric tunnel junctions (FTJs) is the high resistance-area ($RA$) product due to the existence of the critical thickness for the ferroelectric polarization of tunnel barrier [17].

Recently, an alternative route to obtain moderate TMR has been proposed via the magnetic phase transition (MPT) of a magnetic electrode [18,19]. A typical structure of this type of MTJ might be a sandwich structure of "metallic MPT-electrode/insulating tunnel barrier/non-magnetic metallic electrode". The magnetoresistance will arise when the internal magnetic structure of MPT-electrode changes by external magnetic field, temperature, *etc*. For instance, the α'-FeRh electrode can be switched from G-type antiferromagnetic (AFM) to FM state via magnetic field (at the order of several Tesla), and over 20% TMR at room temperature has been demonstrated experimentally in α'-FeRh/MgO/γ-FeRh MTJ [18]. In the present work, we will focus on another more attractive metallic AFM MPT-material $Mn_3Pt$ and discuss the corresponding magnetoresistance effect.

As a well-studied MPT-materials, the ground magnetic state of cubic $Cu_3Au$ type $Mn_3Pt$ is triangular non-collinear AFM (D-phase). It shows a first-order magnetic transition from D-phase to the collinear AFM state (F-phase) when the temperature rises up to $T_{tr}$~365 K [20,21], and it further becomes paramagnetic when the temperature is high over $T_N$~475 K. The magnetic structures of the two stable AFM states in bulk $Mn_3Pt$ are illustrated in Fig. 1(a). The non-collinear D phase state with the magnetic moments of the three Mn atoms in the unit cell establish a triangular arrangement within the (111) plane. The F phase is a collinear magnetic structure with doubled unit cell along the c direction.

What's more interesting, recent experiments demonstrate that being grown on a ferroelectric $BaTiO_3$ substrate, the transition temperature of $Mn_3Pt$ can be shifted upwards by applying electric field through $BaTiO_3$ substrate. Hence, the collinear F-phase of $Mn_3Pt$ can be driven into non-collinear D-phase by applying an electric field on piezoelectric BaTiO$_3$ substrate above room temperature [22]. Therefore, it would be possible to design tunnel junctions by employing such electric field controlled MPT of Mn$_3$Pt as it is shown in Fig. 1(b). On the other hand, AFM materials are believed to be promising for applications in spintronic devices. However, the lack of efficient method for electrical control of AFM states is one of the main obstacle for their applications in magnetic memory devices. Therefore, the possibility of designing the electric field controlled AFM Mn$_3$Pt-based tunnel junctions with large TMR above room temperature may provide new avenue for AFM spintronics [23,24].

Now we will first briefly discuss the mechanism of electric field controlled MPT of Mn$_3$Pt on a piezoelectric substrate. The transition temperature $T_{tr}$ of Mn$_3$Pt between D and F phase has been found to be closely related to its lattice constant [25] due to the so-called "magneto-volume" effect [26]. The smaller lattice constant will result in higher transition temperature $T_{tr}$ of Mn$_3$Pt. When Mn$_3$Pt is grown on a piezoelectric substrate, for instance BaTiO$_3$, the applied electric field across BaTiO$_3$ will lead to the decrease of in-plane lattice constant due to the inverse piezoelectric effect, and correspondingly, the MPT temperature of Mn$_3$Pt will shift to a higher value. In consequence, the electric field can be applied to switch the magnetic phase of Mn$_3$Pt between collinear F phase and non-collinear D phase at certain temperature window as it has been demonstrated experimentally [22].

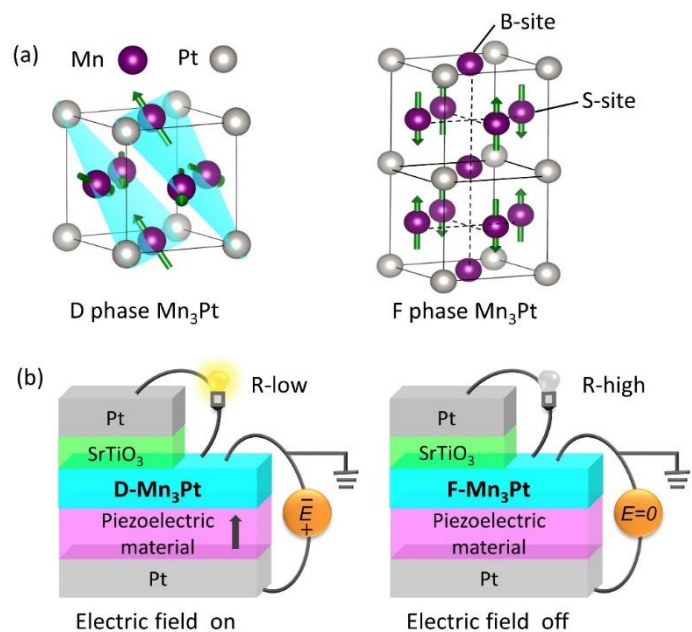

Fig. 1. (a) The illustrations of the atomic and magnetic structures of AFM Mn$_3$Pt in D and F phases. The Pt atoms are in grey and the Mn atoms are in purple. The green arrows indicate the magnetic moment directions on Mn atoms. (b) The demonstration and operation principles of electric field driven MPT-TMR effect in a tunnel junction with a Mn$_3$Pt electrode.

A realistic Mn$_3$Pt-based AFM junction structure and its electric field driven operation principle is shown in Fig. 1(b). The whole Mn$_3$Pt/SrTiO$_3$/Pt tunnel junction can be grown on a piezoelectric substrate, for example, BaTiO$_3$. The electric field is applied between BaTiO$_3$ substrate and Mn$_3$Pt to write the AFM state of Mn$_3$Pt, and the tunnel resistance is measured between Mn$_3$Pt and Pt electrode across the SrTiO$_3$ tunnel barrier. By comparing with conventional MTJs and FTJs, the advantages of the proposed tunnel junctions are multifold including: (1) The magnetic phase of Mn$_3$Pt is controlled by electric field rather than magnetic field or electric current. In consequence, the proposed AFM MTJs may be more energy efficient in magnetoresistance based memory device. (2) The magnetoresistance is originating from the MPT of Mn$_3$Pt electrode and large TMR is optimistic. High TMR of the tunnel junctions will then be beneficial to its application in a memory device with high On/Off ratio. (3) The electric field controlled MPT is switched between two AFM phases, and the whole tunnel junction might be robust against the external magnetic field perturbation and suitable for high density non-volatile memory. (4) The proposed tunnel junction structure is simple with only one magnetic electrode and the switching speed should be fast due to the AFM phase transition.

The remaining key issue of the proposed AFM tunnel junction will then be the spin-dependent transport properties. Especially, the conductance (resistance) ratio of the tunnel junction when Mn$_3$Pt is in F and D-phase which is the main focus of the present work. In the past, there are also several experiments focusing on AFM tunnel junctions by using collinear antiferromagnet L1$_0$-IrMn [27,28] or L1$_0$-MnPt[29] as electrode. However, the magnetoresistances in those tunnel junctions which originates from the anisotropic electronic structure of antiferromagnets are relatively low (typically less than 10%). Hereafter, we will investigate the transport and the corresponding MPT-TMR effect in a Mn$_3$Pt/SrTiO$_3$/Pt tunnel junction through first-principles calculations.

## II. CALCULATION METHODS

Experimentally, a Mn$_3$Pt film originally in the collinear F-phase with a = 3.875 Å, c = 3.850 Å can be switched to the non-collinear D-phase with a = 3.866 Å, c = 3.860 Å above room temperature by applying a moderate out-of-plane electric field E = 4 kVcm$^{-1}$[22]. The corresponding experimental lattice constants of Mn$_3$Pt for two magnetic phases have been used in the present first-principles calculations. All the calculations in the present work are performed by employing the Quantum Espresso package[30] with the PBE-GGA (Perdew-Burke-Ernzerhof type of generalized-gradient-approximation) exchange correlation potential[31] and ultrasoft pseudopotential[32] generated from PSlib0.3.1. A Monkhorst-Pack $k$-point mesh of $16\times16\times16$ and plane-wave cutoff 50 Ry are adopted for the self-consistent electronic structure calculations of bulk Mn$_3$Pt. The calculated ground state of bulk Mn$_3$Pt is D phase which has a total energy 1.01 eV per formula cell lower than the F phase, which agrees with the previous theoretical value [33]. The magnetic moment on Mn atom of D phase is 3.10 $\mu_B$ and it agrees well with the experimental value of 3.0 $\mu_B$[20,21]. For F phase, the Mn spins on the S-sites is 2.92 $\mu_B$ per Mn atom, and on the B-sites is nonzero, but has a small value of 0.11 $\mu_B$ along c axis, resulting a net magnetic moment and magnetization accordingly.

In order to calculate the electron transmission of Mn$_3$Pt/SrTiO$_3$/Pt tunnel junction, first, the electronic structures of the left Mn$_3$Pt electrode, the right Pt electrode, and the junction region in a Mn$_3$Pt/SrTiO$_3$/Pt supercell are separately self-consistently calculated. Then, the electron transmission is calculated by using a standard wave-function scattering method [32,34] in two-dimensional Brillouin zones (2DBZs) by matching the wave function between left and right electrodes. The ballistic Landauer conductance of the tunnel junction is calculated by summarizing the transmission over 200×200 $k_{//}=(k_x, k_y)$ points in 2DBZ: $G=\frac{e^2}{h}\sum_{k_{//}}T(k_{//})$, where $T(k_{//})$ is the $k$-resolved transmission, $e$ is the elementary charge, and $h$ is the Plank constant.

### III. RESULTS AND DISCUSSIONS

The Fermi surface (FS) of electrode indicate the available Bloch states distributed over Brillouin zone for electron transmission, and it is crucial for electron transport of MTJs. Fig. 2(a) shows the side and top view (along $k_z$ direction) of the three-dimensional FSs for bulk Mn$_3$Pt in D and F phases. Both AFM phases have multiple bands and distributed over the 2DBZ, and the noteworthy difference of the two AFM phases is that there is no available Bloch state around the zone center for F phase while several Bloch states are present for D phase, which is evident from the top views shown in Fig. 2(a). This is one of the main reasons for the different transmission in Mn$_3$Pt-based MTJs and the resultant large magnetoresistance through SrTiO$_3$ barrier as we discuss later. The corresponding density of states for D phase and F phase are shown in Fig. 2(b). At Fermi energy, the F-phase has relative larger density of states than D-phase Mn$_3$Pt.

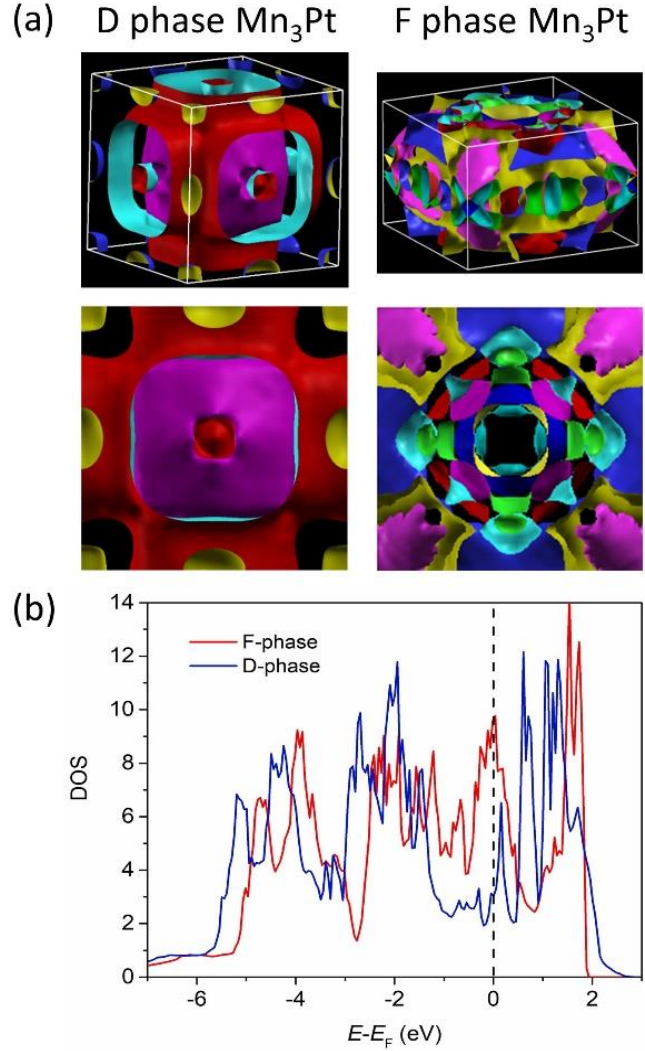

Fig. 2. (a) The side and top views of three-dimensional FSs for Mn$_3$Pt in D and F phases. The FSs are visualized by using Xcrysden package [35]. (b) Density of states (DOS) of bulk Mn$_3$Pt per formula cell for F (red) and D (blue) phases. The Fermi energy lies at zero as indicated by the vertical black dash line.

A Mn$_3$Pt/SiTrO$_3$/Pt MTJ is built and shown in Fig. 3(a). The tunnel junction consists of a semi-infinite Mn$_3$Pt electrode and SrTiO$_3$ tunnel barrier with a thickness of 2 unit cells (u.c.) stacked along the [001] direction. Perovskite oxide SrTiO$_3$ with lattice constant 3.905 Å is chosen as the tunnel barrier by considering its small lattice mismatch (< 1.0 %) with Mn$_3$Pt. In addition, Pt with lattice constant of 3.916 Å[36] serves as counter electrode which receives tunneling electrons from Mn$_3$Pt electrode. Therefore, the whole Mn$_3$Pt/SrTiO$_3$/Pt tunnel junction may be grown epitaxially with small lattice mismatch. The epitaxial relation between Mn$_3$Pt and SrTiO$_3$ might be Mn$_3$Pt(100)[001]// SrTiO$_3$(100)[001]// Pt[001] with the Ti atoms sit at the top of the Mn atoms. In the junction, the left lead consists of repeating unit cells of Mn₃Pt and terminates on both ends with a Mn-Pt atomic layer. The SrTiO₃ layer is terminated on both sides with TiO₂ atomic plane. The oxygen atoms are connected with Mn and Pt atoms at the interface which is energy favorable [16].

The main results of the transport calculation are shown in Fig. 3(b) which displays the $k_{//}$-resolved transmission for the junction with Mn₃Pt electrode in D and F phases. In the junction with D phase Mn₃Pt, an area has the largest transmission of $10^{-1}$ distributed around the 2DBZ center. In contrast, the FS of Mn₃Pt in F phase viewed along the [001] direction has holes in the zone center. There are no bulk states in both spin channels of F phase Mn₃Pt, which results in zero transmission around this area. Accordingly, the total transmission of the Mn₃Pt/SrTiO₃/Pt junction with D phase Mn₃Pt electrode is relatively larger than that with the F phase Mn₃Pt electrode. When the electric field is applied across piezoelectric substrate, the Mn₃Pt electrode undergoes a phase transition from F phase to D phase, consequently the tunneling junction undergoes a transition from high-resistance state to low-resistance state. The optimistic MPT-TMR can be defined as: $\text{MPT-TMR} = \frac{G_{\text{D-phase}} - G_{\text{F-phase}}}{\min(G_{\text{F-phase}}, G_{\text{D-phase}})} \times 100\%$. The electron transmission and the corresponding MPT-TMR are listed in Table 1. One can see that the MPT-TMR is over 500% by this definition.

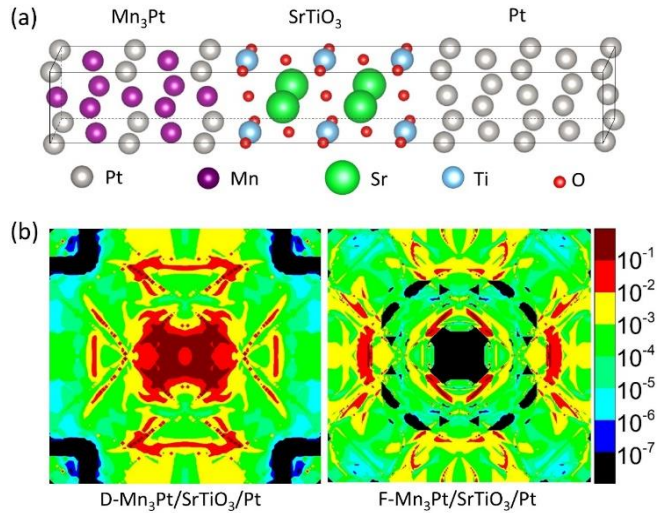

Fig. 3. (a) The side view of the atomic structure of Mn₃Pt/SrTiO₃/Pt tunnel junction. (b) The electron transmission of Mn₃Pt/SrTiO₃/Pt MTJ with Mn₃Pt in D (left) and F (right) AFM phases, respectively. The color bar shows the intensity of transmission.

Table 1. The electron transmission of D-Mn₃Pt and F-Mn₃Pt per formula cell[a]. The transmission of MTJ with paramagnetic Mn₃Pt (P-Mn₃Pt) electrode has also been listed for comparison.

|  | Transmission for D-Mn$_3$Pt | Transmission for F-Mn$_3$Pt | Transmission for P-Mn$_3$Pt | MPT-TMR |
|---|---|---|---|---|
| Bulk-Mn$_3$Pt | 0.85 | 1.87 | — | -120% |
| Mn$_3$Pt/SrTiO$_3$/Pt | 2.5×10$^{-2}$ | 0.40×10$^{-2}$ | 0.90×10$^{-2}$ | 525% |
| Mn$_3$Pt/AlO$_x$/Pt | — | — | — | -200%[b] |

[a]For F-Mn$_3$Pt the transmissions for two spin channel are identical and for non-collinear D-Mn$_3$Pt the transmissions of the two spin channels are indistinguishable.

[b]Estimated from density of states at Fermi energy.

The large MPT-TMR in Mn$_3$Pt/SrTiO$_3$/Pt tunnel junction also partly relies on symmetry selective filtering effect in SrTiO$_3$ tunnel barrier. As it is shown in Fig. 4(a), similar to the band symmetry filtering effect in MgO[1] and spinel oxide MgAl$_2$O$_4$[37], the Bloch state with $\Delta_1$ and $\Delta_5$ symmetry has smallest decay rate within the bandgap of SrTiO$_3$. In consequence, the Bloch states around Γ point (zone center) may have relatively larger tunneling possibility. Fig. 4(b) shows complex wave vector with smallest imaginary part over the entire 2DBZ. It is clear that the slowest electron decay rate forms a cross shape around the Γ point. By comparing the FSs of D-Mn$_3$Pt and F-Mn$_3$Pt shown in Fig. 2(a), there is no available Bloch states of F-Mn$_3$Pt around Γ point while available Bloch states present for D-Mn$_3$Pt. In consequence, for D-Mn$_3$Pt/SrTiO$_3$/Pt tunnel junction, because of large transmission contribution from the Brillion zone center, the tunneling conductance should be much larger than that of F-Mn$_3$Pt/SrTiO$_3$/Pt MTJ, and lead to large positive MPT-TMR.

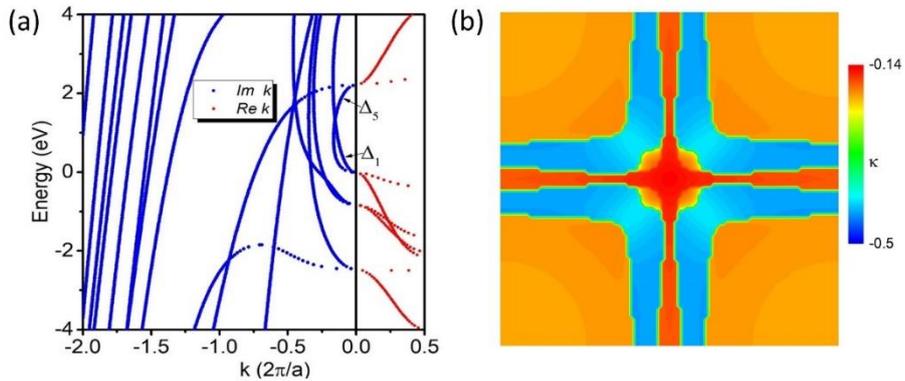

Fig. 4. (a) The complex band structure of SrTiO$_3$ along [001] direction ($k_\parallel$=0). The real band and the imaginary band are plotted in red and blue, respectively. The top of the valence band is located at the zero energy. (b) The two dimensional complex band dispersion of SrTiO$_3$ at the middle of bandgap with smallest imaginary part κ (in the unit of 2π/a).

It is worthwhile pointing out that SrTiO₃ may not be the unique tunnel barrier material suitable for the proposed tunnel junctions with Mn₃Pt electrode. Despite of the large MPT-TMR in the single-crystalline tunnel junction with SrTiO₃ barrier, it is also possible to fabricate a similar junction but with an amorphous barrier, for instance, AlO$_x$ barrier. By ignoring the difference of tunneling ability of each Bloch state, the electron transmission (or conductance) at zero bias is proportional to the density of states at Fermi energy of two electrodes as: $T \propto D_{Pt}(E_F)D_{Mn_3Pt}(E_F)$. A simple estimation of the MPT-TMR with amorphous AlO$_x$ barrier according to the density of states of F-Mn₃Pt and D-Mn₃Pt shown in Fig. 2(b) is around -200 % as it is listed in Table 1. Here the negative sign indicates a larger conductance of F-Mn₃Pt/AlO$_x$/Pt than D-Mn₃Pt/AlO$_x$/Pt MTJ.

The interfacial magnetic structure may be important for the spin-dependent transport and the resultant TMR in a MTJ [38]. In order to elucidate the effect of possible interface magnetic disorder, additional calculations are performed. Take the F-Mn₃Pt/SrTiO₃/Pt-MTJ for example, instead of perfect AFM interface magnetic structure (shown in Fig.5 (a)), we consider 1 u.c. (shown in Fig. 5(b)) and 4 u.c. (not shown) of paramagnetic Mn₃Pt present at the interface. The resultant electron transmission is listed in Table 2. Comparing with the perfect AFM ordered Mn₃Pt interface, paramagnetic Mn₃Pt present at the interface will lead to additional interface scatting and decrease of electron transmission. However, the MPT-TMR has been largely preserved due to the fact that the MPT-TMR mostly originates from the features of FS of bulk Mn₃Pt as we discuss previously. These results further confirm that the MPT-TMR should be robust against the imperfect interfacial magnetic structure.

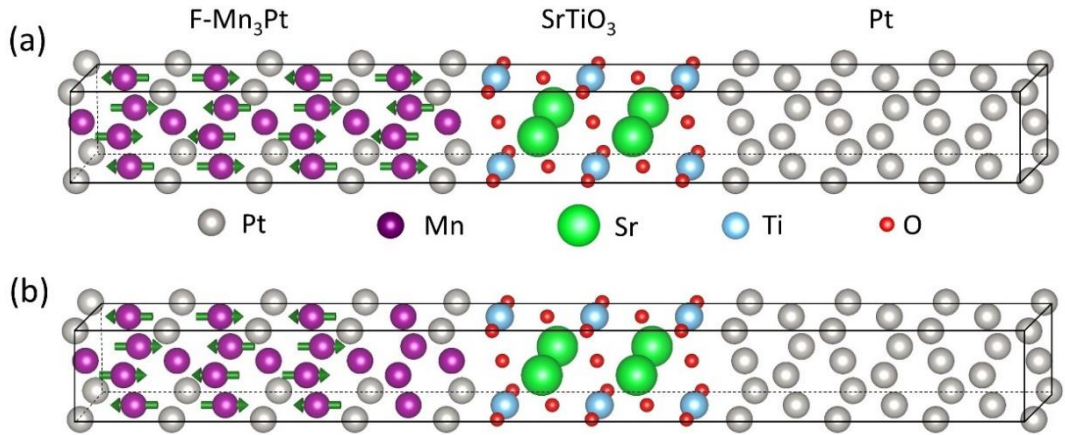

Fig. 5 Atomic structure of F-Mn₃Pt/SrTiO₃/Pt MTJ with perfect interface magnetic structure (a) and with 1 u.c. of paramagnetic Mn₃Pt interface (b). The arrows represent the magnetic moments.

Table 2. The electron transmission of F-Mn₃Pt/SrTiO₃/Pt-MTJ per formula cell with different

interfacial magnetic structures and the corresponding MPT-TMR respecting to D-Mn$_3$Pt/SrTiO$_3$/Pt-MTJ.

| Interfacial magnetic structure | Transmission | MPT-TMR |
|---|---|---|
| 1 u.c. P-Mn$_3$Pt | 0.11×10$^{-2}$ | +2173% |
| 4 u.c. P-Mn$_3$Pt | 0.28×10$^{-2}$ | +793% |

In stoichiometry Mn$_3$Pt/BaTiO$_3$ system, the electric field control of MPT occurs above room temperature at around 360 K [22]. It may also be possible to further reduce the MPT temperature to room temperature (300 K) by choosing appropriate materials. For instance, it has been experimentally shown that the MPT temperature of non-stoichiometry Mn$_3$PtN$_x$ and Mn$_{3-x}$Pt$_{1+x}$ alloys [25,39,40] can be lowered down to room temperature, and theoretically shown that Mn-based antiperovskite nitrides [41] may have similar AFM phase transition. And also instead of BaTiO$_3$ substrate, an alternative piezoelectric substrate with a larger piezoelectric effect, for example, PMN-PT (Pb(Mg$_{1/3}$Nb$_{2/3}$)O$_3$-PbTiO$_3$) [42] substrate may lead to a MPT at room temperature.

Moreover, the electronic structures of Mn$_3$Pt in the non-collinear AFM D-phase and collinear AFM F-phase are significantly different and may lead to other different intrinsic physical properties beyond magnetoresistance effect in the studied Mn$_3$Pt-based MTJs. For example, it has been experimentally demonstrated that the non-collinear AFM D-phase has moderate anomalous Hall conductivity (AHE) while collinear AFM F-phase has zero AHE conductivity [22]. Thus one may expect different spin hall conductivity (or spin-hall angle) of Mn$_3$Pt in two different AFM phases [43]. In addition, non-collinear and collinear AFM Mn$_3$Pt may lead to different exchange bias effect in Mn$_3$Pt/FM metal bilayer system. All these physical properties, can be controlled by electric field through inverse piezoelectric effect [22,44,45] and used in AFM spintronics [23,24].

## IV. SUMMARY

In summary, by employing first-principles calculations we have investigated the transport properties of MTJs by using an AFM electrode Mn$_3$Pt, which can be transformed from collinear AFM F-phase to non-collinear AFM D-phase by applying electric field across piezoelectric substrate. Our results show that the magnetoresistance ratio in a Mn$_3$Pt/SrTiO$_3$/Pt tunnel junction can reach hundreds of percent, making it promising for application in low-power consumption memory devices. The MPT-TMR in the proposed single crystalline MTJ originates from the cooperative effect of different Fermi surfaces of

Mn$_3$Pt in two AFM phases and the band symmetry filtering effect of SrTiO$_3$ barrier. This would make the MPT-TMR robust against the possible interface magnetic structure disorder. In addition, by estimating from the density of states, a similar tunnel junction with amorphous AlO$_x$ barrier also has large MPT-TMR. Such electric field controlled MPT-TMR effect is expected in a class of similar materials beyond Mn$_3$Pt. Moreover, the electric-field controlled MPT of Mn$_3$Pt can be largely extended to other electric-controlled phenomenon beyond MPT-TMR including anomalous Hall, spin Hall, exchange bias, *etc.* This work may stimulate the future experimental investigations on the MPT-TMR mechanism and the application of Mn$_3$Pt and similar materials in AFM spintronics.

## ACKNOWLEDGMENTS

Jia Zhang and Long You are supported by the National Natural Science Foundation of China with grant No. 11704135, 61674062 and 61821003. The calculations in this work are partly performed at National Supercomputer Center in Tianjin, TianHe-1(A) China.